\documentclass[onecolumn,preprintnumbers,amsmath,amssymb,superscriptaddress]{nature}
\usepackage{amssymb}
\usepackage{mathrsfs}
\usepackage{graphicx}
\usepackage{dcolumn}
\usepackage{bm}
\usepackage{amsmath}
\usepackage{amsfonts}
\usepackage{color}
\usepackage{float}

\title{Two dimensional photonic quasicrystal edge states protected by second Chern number}

\author{Xiao Zhang$^1$}

\begin{document}

\maketitle

\begin{affiliations}
 \item Department of Physics, Sun-Yat-Sen University, Guangzhou, China.
Correspondence and requests for materials should be addressed to Xiao Zhang (email: yngweiz@gmail.com)
\end{affiliations}

\begin{abstract}
\textbf{Topological physics in photonic systems have attracted great attentions in recent years. In this work, we theoretically study the one and two dimensional photonic quasicrystal resonator lattices characterized by the first and second Chern number, which show exotic boundary states within the photonic energy band gap. In particular, the second Chern number protected edge states has opened up new possibilities for realizing topological physics of dimensions higher than three in photonic systems, which is highly sought for. Such photonic systems can be easily experimentally realized in
regular photonic crystal with dielectric rods in air, by varying the radius of the rods, so we propose experiments realizing our predictions.}
\end{abstract}%much more exotic though previously neglected nonlinear electromagnetic response of material systems with a Dirac Ring, which has been The characteristic response of a Dirac Ring }

\date{\today}

\maketitle
%\tableofcontents

\section{Introduction}
 Topological phases in condensed matter including quantum Hall effect (QHE)$^{1}$, topological insulators$^{2-7}$, and quantum anomalous Hall effect (QAHE)$^{8-9}$ have recently garnered tremendous interest in
condensed-matter physics, material science and electrical
engineering communities$^{10}$. In those systems, each energy gap is attributed an topological index. A nontrivial topological 
index is usually associated with interesting boundary states,
quantized response, and exotic quasiparticle excitations. Soon after the discovery of these topological phases in condensed matter, similar topological effects in photonic systems $^{11-12}$ have been proposed and demonstrated, in analogy to the topological physics in condensed matter. Although topological phases exist in any dimension, those systems have dimensions below 3D, the maximum real space dimension in our universe. Thus for topological phases described by Chern number, only 2D systems protected by the first Chern number including QHE, QAHE and its photonic analogy $^{11-12}$ are realized. Topological phases protected by the second Chern number exist in 4D$^{13}$, and thus were thought to exist in theory only.  

Only recently, using the physical properties of quasicrystals (QCs)-
nonperiodic structures with long-range order, 4D QHE protected by second Chern number have been proposed in 2D photonic systems$^{14}$. This is due to QCs can
oftentimes be derived from periodic models of higher
dimensions, so they contain extra information in higher dimensions$^{14-17}$. The essential physics in their proposal is the on-site modulation of energy $^{14-15}$. However, in this proposal, $z$ dimension is treated as time and thus the wave vector in $z$, $\beta_z$ represents energy. This is different than the conventional photonic crystal (PC) context, in which the frequency $\omega$ is the energy and its experimental setup has been very mature to realize exotic photonic physics.

It is known that the resonant frequency of resonators in photonic crystals can be modulated through their size$^{18}$. With this principle, in this work we propose a 2D photonic crystal resonator lattice that have second Chern number protected edge states within the band gap of frequency $\omega$ . The proposed structure can be easily manufactured with modern fabrication technology.  
\section{Results}
We first introduce some basic theory on the tight binding formalism of resonators in PCs. With that, we  first illustrate our idea with first Chern number protected one dimensional PC resonator boundary states, then present our main results on the second Chern number protected two dimensional PC resonator boundary states.
\subsection{Resonator Defect States in Photonic Crystal}
\label{rl}
\begin{figure}[H]
\includegraphics[scale=.7]{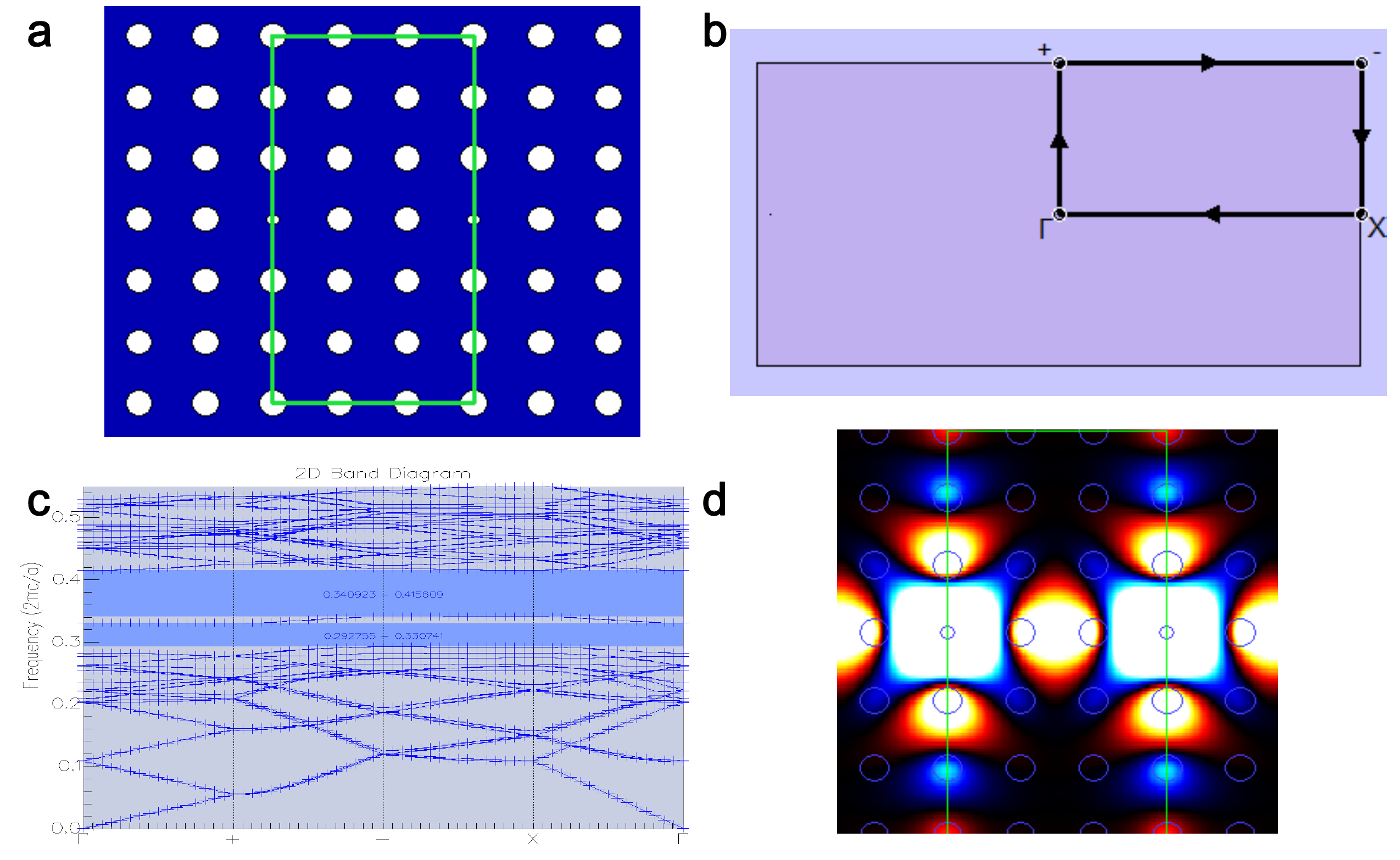}

\caption{(Color online) (a) The one dimensional resonator array (small white rods) embedded in two  dimensional photonic crystal of  dielectric
rods (big white rods) with radius 0.20a and refractive index 3.4 (AlGaAs). The resonator defect is created by changing the
radius of a single rod. For example, the case where $r=0$ corresponds to the removal
of a rod. The green box indicates the unit cell we use to simulation the energy band diagram of the whole structure, including both the 2D photonic crystal and its embedded defect array.  (b)The Brillouin zone of the unit cell in (a). (c) The energy band diagram of the whole unit cell in (a) in TM mode, the dielectric
rods (big white rods) with radius 0.20a form a band structure with band gap $0.29-0.42(2\pi c/a)$, inside the band gap are the tight binding states of the 1D resonator defect array with radius $0.0965a$.(d) The electric field distribution of the defect states in TM mode. The defect states are monopole states and form a tight binding lattice.}
\label{fig:1}
\end{figure}

In this work, we consider the square lattice PC  composed of dielectric
rods with radius 0.20a and refractive index 3.4 in air (Fig. \ref{fig:1}a)$^{18}$. Its TM modes are well known to have a band gap from $0.29-0.42(2\pi c/a)$$^{18}$. Like dopant in a semiconductor, when the radius of a rod varies from 0.20a, the defect states form localized states in the photonic band gap. Using a hamiltonian formalism$^{19}$, where the resonator mode $\left|\psi\right>=\left[
\begin{array}[]{c}
\vec E\\
\vec H\\	
\end{array}\right]$
is the eigenvector and the resonant 
frequency $\omega$ is the eigenvalue of the equation, we have
\begin{equation}
\Theta_0 \left|\psi\right>=\omega \left|\psi\right>
\label{H0}
\end{equation}
with 
\begin{equation}
\Theta_0= \left[
\begin{array}[]{cc}
0 & i\varepsilon_{0}^{-1}\vec\nabla\times\\
-i\mu_0^{-1}\vec\nabla\times & 0\\	
\end{array}\right]
\label{H0exp}
\end{equation}
Here $\omega$ is the resonant frequency lying within the bulk PC band gap (Fig. \ref{fig:1}c). $\varepsilon_0(x,y)$ is a function of the refractive index in real space. Inside the dielectric rods $\varepsilon_0=3.4$, outside  $\varepsilon_0=1$. When the radius of the defect $r<0.2a$, the defect states are monopole states$^{18}$ (Fig.\ref{fig:1}d).

\subsection{One Dimensional Photonic Crystal Resonator Lattice}
\label{1d}
When we have more than one resonators that are close enough in space, they are no longer isolated and the perturbed Hamiltonian is now
$\Theta=\Theta_0+V$, where
\begin{equation}
V= \left[
\begin{array}[]{cc}
0 & i(\varepsilon^{-1}-\varepsilon_{0}^{-1})\vec\nabla\times\\
0 & 0\\	
\end{array}\right]
\label{H0exp}
\end{equation}
The perturbation terms result in 
\begin{equation}
V_{\alpha\beta}=\frac{-\int \Delta\varepsilon\omega_0 E_{\alpha}^*E_{\beta}dV}{2\sqrt{\int\varepsilon_0\left|E_{\alpha}\right|^2dV}\sqrt{\int\varepsilon_0\left|E_{\beta}\right|^2dV}} 
\label{V}
\end{equation}
where $\Delta\varepsilon=\varepsilon-\varepsilon_0$ is the difference of spatial dependent
refractive index of the unperturbed system (single defect) $\varepsilon_0$ and the perturbed system (coupled defects) $\varepsilon$$^{20}$.
$V_{\alpha\alpha}$ is the self-energy correction for $\omega_0$ of defect $\alpha$$^{21}$, and $V_{\alpha\beta}$ is the hopping between defect $\alpha$ and $\beta$$^{20}$. Defining $\omega_{\alpha}=\omega_0+V_{\alpha\alpha}$, we can write the hamiltonian as 
\begin{equation}
H=\sum_{\alpha}\omega_{\alpha}\left|\psi_{\alpha}\right>\left<\psi_{\alpha}\right|+\sum_{\alpha,\beta}V_{\alpha\beta}\left|\psi_{\alpha}\right>\left<\psi_{\beta}\right| 
\label{H1}
\end{equation}
For an 1D resonator lattice of lattice constant $3a$ with translational symmetry (Fig. \ref{fig:1}a), considering nearest neighborhood hopping only, $V_{\alpha,\alpha+1}=V_{\alpha-1,\alpha}=t$, $\omega_{\alpha}=\omega_c$. Using Bloch theorem, we can solve the energy to be 
\begin{equation}
\omega(k)=\omega_c+2t\cos(3ka)
\label{E1}
\end{equation}
as shown in Fig. \ref{fig:1}c. By fitting Eq. \ref{E1}, we can get for $r=0.0965a$, $\omega_c=0.3312231(2\pi c/a)$, $t=-0.00485(2\pi c/a)$. Fig. \ref{fig:2}a shows the $w_c$ for lattice of resonators when we change the radius. 
\subsection{One Dimensional Photonic Quasicrystal Resonator Lattice}
\label{1dQC}
We have discussed 1D photonic crystal resonator lattice composed of resonators with translational symmetry. Next we will elaborate how to construct a quasicrystal resonator lattice with bulk bands characterized by the first Chern number. Here the word ``quasicrystal'' does not mean a variation of lattice spacing, but a variation of each resonator's radius. The variation of resonator's size introduces the on-site modulation of energy needed for nontrivial Chern number$^{14-15}$. Here are the construction steps:

(1)  Using $\omega_{c0}$ of $r=0.0965a$ as the zero energy reference, we compute the on-site energy of resonator site $x$: $\omega_x=\omega_{c0}+\lambda_x\cos(2\pi bx+\phi_x)$. $b$ is an irrational number. In this paper, we use $b=(1+\sqrt{5})/2$. $x$ is an integer index for sites.

\begin{figure}[H]
\includegraphics[scale=.7]{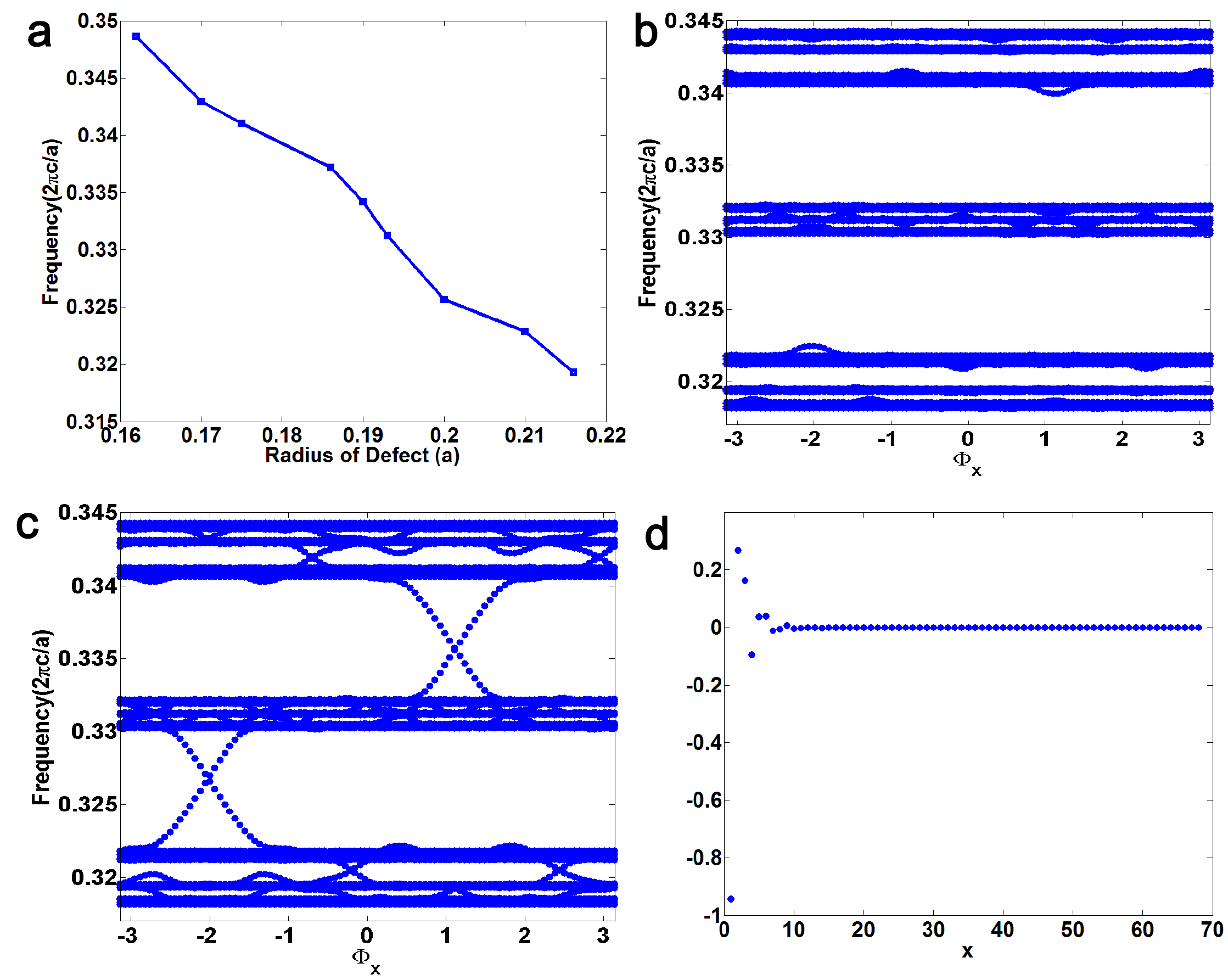}
\caption{(Color online) (a)The energy at $\Gamma$ point (Fig. \ref{fig:1}c) of the 1D resonator defect lattice of various sizes embedded in an 2D PC of dielectric
rods with radius 0.20a (Fig. \ref{fig:1}a). (b) The bands of the 1D defect QC (Fig. \ref{fig:1}a)constructed in Section \ref{1dQC} with periodical boundary condition according to Eq. \ref{H1c}, the number of sites are chosen to be 68, and the parameters are $t=-0.00485$,
$\lambda=0.010412$, $b=(1+\sqrt{5})/2$.(c) The bands of the same 1D resonator QC as in (b) but with open boundary condition instead. (d) From (c), we can observe that when $\phi_x=1.174$ and $\omega_x=0.3361(2\pi c/a)$, only boundary state exists. Here we plot the wave function of this state using the tight binding basis at different sites $x$.}  
\label{fig:2}
\end{figure}

(2) Using the a monotonic function of $\omega_x$ of the resonator radius in  1D resonator lattice with translational symmetry discussed in last section as shown in Fig. \ref{fig:2}a, we 
can determine the radius $r_x$ for resonator $x$ by inverting the function.

Now we get a resonator QC lattice, each site with a different size. Notice that because $\omega_x$ in step (2) was computed assuming 1D translational symmetry, now for the quasicrystal lattice without translational symmetry, $\omega_x$ is modulated by its neighborhood with different sizes, same for the hopping parameter $t$.   
Considering nearest neighborhood hopping only, now the hamiltonian is 
\begin{equation}
H=\sum_{x}\omega_{x}\left|\psi_{x}\right>\left<\psi_{x}\right|+\sum_{x,x+1}V_{x,x+1}\left|\psi_{x}\right>\left<\psi_{x+1}\right| +H.c.
\label{H1a}
\end{equation}
whereas we have defined functions $\omega_x=\omega((2\pi bx+\phi_x)mod2\pi)$, $V_{x,x+1}=V_1((2\pi bx+\phi_x)mod2\pi)$, $V_{x,x-1}=V_2((2\pi bx+\phi_x)mod2\pi)$. These functions can be defined because according to our steps to construct the lattice, $(2\pi bx+\phi_x)mod2\pi$ determines the size of the resonator at $x$ as well as its neighborhood. Next we shall proceed to show Eq. \ref{H1a} describe a Chern insulator with first Chern number. 

In order for the Chern number to be defined, we introduce a twisted boundary condition by parameter $\theta$$^{14,15}$
\begin{equation}
H=\sum_{x}\omega_{x}\left|\psi_{x}\right>\left<\psi_{x}\right|+\sum_{x,x+1}V_{x,x+1}e^{-i\theta/L}\left|\psi_{x}\right>\left<\psi_{x+1}\right| +\sum_{x,x-1}V_{x,x-1}e^{i\theta/L}\left|\psi_{x}\right>\left<\psi_{x-1}\right|
\label{H1b}
\end{equation}
Next we set $\overline{b}_L=\left\lfloor b\cdot L\rfloor\right/L$ as a rational approximation to b so that the
modulation is periodic. When $L\rightarrow+\infty$ the results will be the same as irrational $b$. When $L$ is a prime number, for a shift $\phi\rightarrow\phi+\varepsilon$ where $\varepsilon=2\pi l/L$ and $l=0, 1, ..., L-1$, we can always find a translation of lattice site $n\rightarrow n+n_{\varepsilon}$, $n_{\varepsilon}\in 0,1,...,L-1$, so that any function $V((2\pi\overline{b}_Ln+\phi+\varepsilon) mod 2\pi)=V((2\pi\overline{b}_L(n+n_{\varepsilon})+\phi)mod 2\pi)$. We denote the translation operator by $n_{\varepsilon}$ sites as $T_{\varepsilon}$, then $H_b(\phi+\varepsilon)=T_{\varepsilon}H_b(\phi)T_{\varepsilon}^{-1}$. The equivalence between phase shift and translation means that in the thermodynamical limit $L\rightarrow+\infty$, the shift has no physical consequence such as gap closing and the energy spectrum is independent of $\phi$. To evaluate the Chern number we define the projector operator for states below the band gap as$^{15}$
\begin{equation}
P(\phi,\theta)=\sum_{E_n<E_{gap}}\left|n\right>\left<n\right|
\label{P1}
\end{equation}
$\left|n\right>$ is the state with energy $E_n$ and $E_{gap}$ is any energy within the band gap. Since the projection operator differs from the hamiltonian only by its eigenvalues, we also have $P(\phi+\varepsilon,\theta)=T_{\varepsilon}P(\phi,\theta)T_{\varepsilon}^{-1}$ and $\partial_{\phi}P(\phi+\varepsilon,\theta)=T_{\varepsilon}\partial_{\phi}P(\phi,\theta)T_{\varepsilon}^{-1}$.
The Chern number can be calculated as
\begin{equation}
C_1=\frac{1}{2\pi i}\int_{0}^{2\pi}d\phi d\theta C(\theta,\phi)
\label{C1}
\end{equation}
where 
\begin{equation}
C(\phi,\theta)=Tr\left(P\left[\frac{\partial P}{\partial\phi},\frac{\partial P}{\partial\theta}\right]\right)
\label{C2}
\end{equation}
is the Chern number density$^{15}$. It can be easily proven that
\begin{equation}
C(\phi+\varepsilon,\theta)=C(\phi,\theta)
\label{C3}
\end{equation}
So $C(\phi, \theta)$ has $1/L$ periodicity. Thus
\begin{equation}
\int_{0}^{2\pi}d\phi C(\phi,\theta)=L\int_{0}^{2\pi/L}d\phi C(\phi,\theta)=2\pi C(\phi=0,\theta)+O(1/L)
\label{C4}
\end{equation}
Next we need to prove $C(\phi,\theta)$ is also independent of $\theta$. From Eq. \ref{H1b}, $e^{i\theta/L}\approx 1+i\theta/L$.  According to perturbation theory, $H$ and $P$ are suppressed by a factor of $1/L$, thus in the thermal dynamic limit, $C(\phi,\theta)$ is independent of $\theta$ as well, and 
\begin{equation}
C1=-(2\pi i) C(\phi=0, \theta=0)+O(1/L)
\label{C4}
\end{equation}
Eq. \ref{H1a} can be well approximated as (See Methods Section) 
\begin{equation}
H=\left(\omega_{c0}+\lambda_x\cos(2\pi bx+\phi_x)\right)\left|\psi_{x}\right>\left<\psi_{x}\right|+\sum_{x,x+1}t\left|\psi_{x}\right>\left<\psi_{x+1}\right| +\sum_{x,x-1}t\left|\psi_{x}\right>\left<\psi_{x-1}\right|
\label{H1c}
\end{equation}
The bands of this model with periodical and open boundary condition are plotted in Fig. \ref{fig:2}b and c, with the number of sites chosen to be 68, $t=-0.00485$
and $\lambda=0.010412$. Here according to Fig. \ref{fig:2}a, $\lambda$ is determined by the range of resonator radius in the QC lattice. For the value we use, $r_x$ varies from $0.0875a$ to $0.1065a$. By comparing those two figures, we can clearly observe the boundary states appearing in the open boundary condition, whose number equals to the Chern number difference of the bulks bands across the gap.  In Fig. \ref{fig:2}d we plot the wave function of the boundary state at $\phi_x=1.174$ and $\omega_x=0.3361(2\pi c/a)$ using the tight binding basis at different sites $x$, which can be simply experimentally realized using the construction steps described in this Section.

\subsection{Two Dimensional Photonic Crystal Resonator Lattice}
\label{2d}
\begin{figure}[H]
\includegraphics[scale=.7]{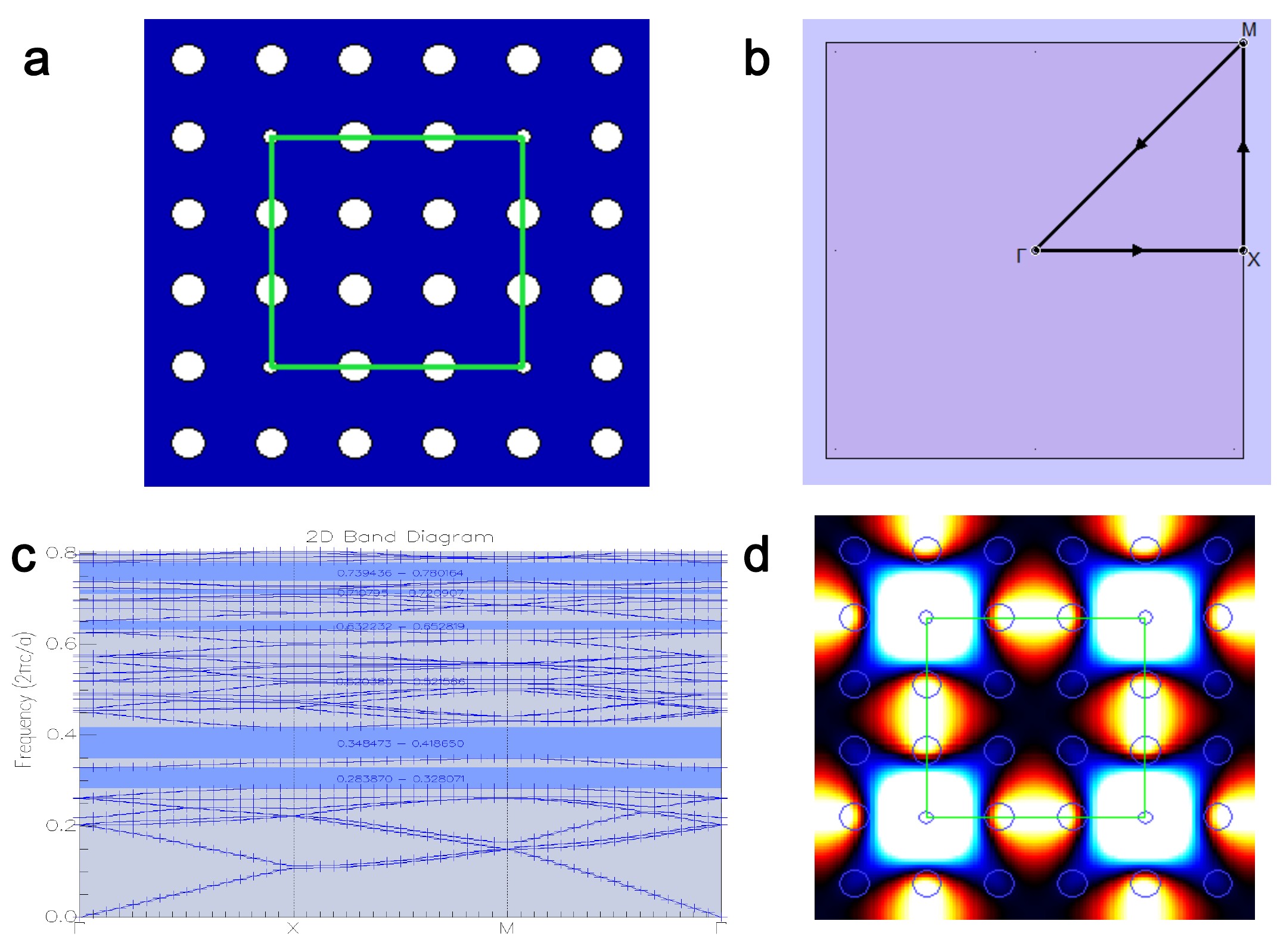}
\caption{(Color online) (a) The two dimensional resonator array (small white rods) embedded in two  dimensional photonic crystal of  dielectric
rods (big white rods) with radius 0.20a and refractive index 3.4 (AlGaAs). The resonator defect is created by changing the
radius of a single rod. For example, the case where $r=0$ corresponds to the removal
of a rod. The green box indicates the unit cell we use to simulation the energy band diagram of the whole structure, including both the 2D photonic crystal and its embedded defect array. (b)The Brillouin zone of the unit cell in (a). (c) The energy band diagram of the whole unit cell in (a) in TM mode, the dielectric
rods (big white rods) with radius 0.20a form a band structure with band gap $0.28-0.42(2\pi c/a)$, inside the band gap are the tight binding states of the 2D resonator defect array with radius $0.095a$. (d) The electric field distribution of defect states in TM mode. The defect states are monopole states and form a tight binding lattice.}
\label{fig:3}
\end{figure}

The hamiltonian to describe two dimensional photonic crystal resonator lattice is still Eq. \ref{H1}, whereas $\alpha=(x,y)$. For a 2D resonator lattice of lattice constant 3a with translational symmetry, considering nearest neighborhood hopping only, $V_{(x,y),(x+1,y)}=V_{(x,y),(x-1,y)}=V_{(x,y),(x,y-1)}=V_{(x,y),(x,y+1)}=t$, $\omega_{(x,y)}=\omega_c$. Using Bloch theorem, we can solve the energy to be 
\begin{equation}
\omega(\vec{k})=\omega_c+2t(\cos(3k_xa)+\cos(3k_ya))
\label{E2}
\end{equation}
as shown in Fig. \ref{fig:3}c. By fitting Eq. \ref{E2}, we can get for $r=0.095a$, $\omega_c=0.3280711(2\pi c/a)$, $t=-0.0051005(2\pi c/a)$. Fig. \ref{fig:4}a shows the $w_c$ for lattice of resonators with different radius. 
\subsection{Two Dimensional Photonic Quasicrystal Resonator Lattice}
\label{2dQC}
We have discussed 2D photonic crystal resonator lattice
composed of resonators with translational symmetry. Next we will present the main results of our paper on how to construct a QC resonator lattice
with bulk bands characterized by the second Chern number in a  way similar to the 1D QC case. The variation of resonator’s size introduces the on-site modulation of energy
needed for a nontrivial Chern number$^{14}$. Here are the construction steps:

(1)  Using $\omega_{c0}$ of $r=0.095a$ as the zero energy reference, we compute the on-site energy of resonator site $(x,y)$: $\omega_{(x,y)}=\omega_{c0}+\lambda_x\cos(2\pi bx+\phi_x)+\lambda_y\cos(2\pi by+\phi_y)$. $b$ is an irrational number. In this paper, we use $b=(1+\sqrt{5})/2$. $x$ is an integer index for sites.

\begin{figure}[H]
\includegraphics[scale=.7]{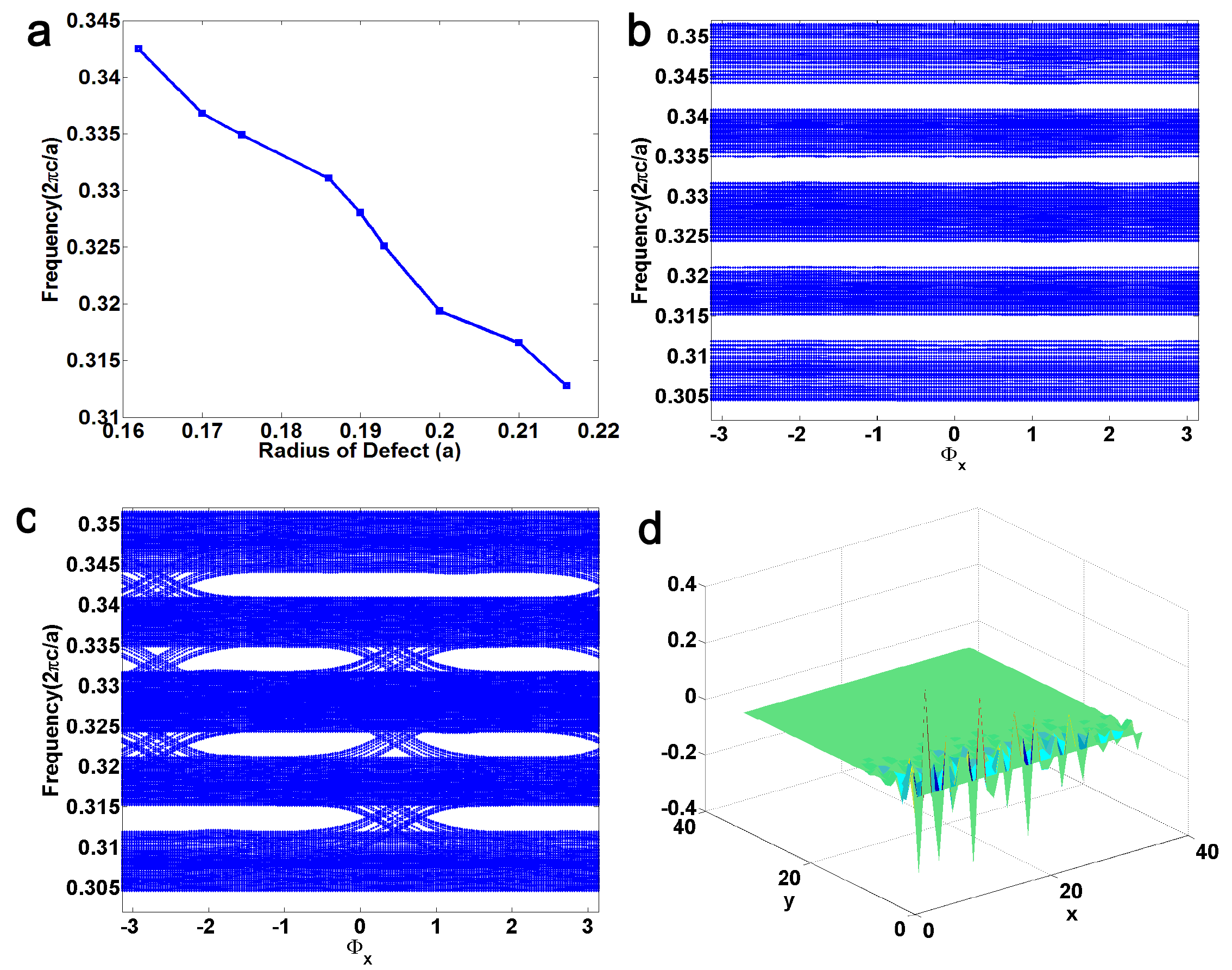}
\caption{(Color online) (a)The energy at $\Gamma$ point (Fig. \ref{fig:3}d) of the 2D resonator lattice of various sizes embeded in an 2D PC of dielectric
rods with radius 0.20a (Fig. \ref{fig:3}a). (b) The bands of the 2D defect QC (Fig. \ref{fig:3}a)constructed in Section \ref{2dQC} with periodical boundary condition according to Eq. \ref{H2c}, the number of sites in both x and y directions are chosen to be 34, and the parameters are $t_x=t_y=-0.0051$,
$\lambda_x=\lambda_y=0.00741925$, $b=(1+\sqrt{5})/2$, $\phi_y=0$.(c) The bands of the 2D defect quasicrystal the same as in (b) but with open boundary condition. (d) From (c), we can observe that when $\phi_x=-2.443$, $\phi_y=0$ and $\omega_x=0.3427(2\pi c/a)$, only boundary state exists. Here we plot the wave function of this state using the tight binding basis at different sites $(x,y)$. The boundary state is localized on one edge.}
\label{fig:4}
\end{figure}

(2) Using the a monotonic function of $\omega_{(x,y)}$ of the resonator radius in the 2D resonator lattice with translational symmetry discussed in last section as shown in Fig. \ref{fig:4}a, we 
can determine the radius $r_{(x,y)}$ for resonator $(x,y)$ by inverting the function.

Now we get a resonator QC lattice, each site with a different size. Notice that because $\omega_{(x,y)}$ in step (2) was computed assuming 2D translational symmetry, now for the QC lattice without translational symmetry, $\omega_{(x,y)}$ is modulated by its neighborhood with different sizes, same for the hopping parameter $t$.   
Considering nearest neighborhood hopping only, now the hamiltonian is 
\begin{equation}
H=\sum_{x,y}\omega_{(x,y)}\left|\psi_{(x,y)}\right>\left<\psi_{(x,y)}\right|+\sum_{x,y}V_{(x,y),(x\pm 1,y)}\left|\psi_{(x,y)}\right>\left<\psi_{(x\pm 1,y)}\right|+\sum_{x,y}V_{(x,y),(x,y\pm 1)}\left|\psi_{(x,y)}\right>\left<\psi_{(x,y\pm 1)}\right| 
\label{H2a}
\end{equation}
whereas we have defined functions $\omega_{(x,y)}=\omega((2\pi bx+\phi_x)mod2\pi,(2\pi by+\phi_y)mod2\pi)$, $V_{(x,y),(x+1,y)}=V_1((2\pi bx+\phi_x)mod2\pi,(2\pi by+\phi_y)mod2\pi)$, $V_{(x,y),(x-1,y)}=V_2((2\pi bx+\phi_x)mod2\pi,(2\pi by+\phi_y)mod2\pi)$, $V_{(x,y),(x,y+1)}=V_3((2\pi bx+\phi_x)mod2\pi,(2\pi by+\phi_y)mod2\pi)$, $V_{(x,y),(x,y-1)}=V_4((2\pi bx+\phi_x)mod2\pi,(2\pi by+\phi_y)mod2\pi)$. These functions can be defined because according to our steps to construct the lattice, $(2\pi bx+\phi_x)mod2\pi$ and $(2\pi by+\phi_y)mod2\pi$ ubiquitously determine the size of the resonator at $(x,y)$ as well as its neighborhood. Next we will elaborate how Eq. \ref{H2b} describes a Chern insulator with second Chern number following the 1D QC case. 

In order for the Chern number to be defined, we introduce a twisted boundary condition by parameter $\theta$$^{14}$
\begin{eqnarray}
H&=&\sum_{x,y}\omega_{(x,y)}\left|\psi_{(x,y)}\right>\left<\psi_{(x,y)}\right|+\sum_{x,y}V_{(x,y),(x+1,y)}e^{-i\theta_x/L}\left|\psi_{(x,y)}\right>\left<\psi_{(x+1,y)}\right| +\sum_{x,y}V_{(x,y),(x-1,y)}e^{i\theta_x/L}\left|\psi_{(x,y)}\right>\left<\psi_{(x-1,y)}\right|\notag\\ 
&+&\sum_{x,y}V_{(x,y),(x,y+1)}e^{-i\theta_y/L}\left|\psi_{(x,y)}\right>\left<\psi_{(x,y+1)}\right| +\sum_{x,y}V_{(x,y),(x,y-1)}e^{i\theta_y/L}\left|\psi_{(x,y)}\right>\left<\psi_{(x,y-1)}\right| 
\label{H2b}
\end{eqnarray}
In the same procedure as in the 1D QC case, by setting $\overline{b}_L=\left\lfloor b\cdot L\rfloor\right/L$ as a rational approximation to b we can prove the equivalence between phase shift and translation in the thermodynamical limit $L\rightarrow+\infty$. So the shift has no physical consequence such as gap closing and the energy spectrum is independent of $\phi$. To evaluate the Chern number we define the projector operator for states below the band gap as 
\begin{equation}
P(\phi_x,\phi_y,\theta_x,\theta_y)=\sum_{E_n<E_{gap}}\left|n\right>\left<n\right|
\label{P2}
\end{equation}
$\left|n\right>$ is the state with energy $E_n$ and $E_{gap}$ is any energy within the band gap. Given $\phi_{\mu}=(\phi_x,\phi_y,\theta_x,\theta_y)$,
the second Chern number can be calculated as
\begin{equation}
C_2=\int_{0}^{2\pi}d\phi_{\mu} C(\phi_{\mu})
\label{C2a}
\end{equation}
where 
\begin{equation}
C(\phi_{\mu})=\sum_{\alpha\beta\gamma\delta}\frac{\varepsilon_{\alpha\beta\gamma\delta}}{-8\pi^2}Tr\left(P\frac{\partial P}{\partial\phi_{\alpha}}\frac{\partial }{\partial\phi_{\beta}}P\frac{\partial P}{\partial\phi_{\gamma}}\frac{\partial P}{\partial\phi_{\delta}}\right)
\label{C2b}
\end{equation}
is the Chern number density. As in the 1D quasicrystal case,
\begin{equation}
C(\phi+\varepsilon,\theta)=C(\phi,\theta)
\label{C2c}
\end{equation}
and $C(\phi,\theta)$ is also independent of $\theta$ in the thermal dynamic limit.
\begin{equation}
C_2=(2\pi)^4 C(\phi_{\mu})\neq 0
\label{C2d}
\end{equation}
Eq. \ref{H2a} can be well approximated as (See Method Section) 
\begin{eqnarray}
H&=&\sum_{x,y}\left(\omega_{c0}+\lambda_x\cos(2\pi bx+\phi_x)+\lambda_y\cos(2\pi by+\phi_y)\right)\left|\psi_{(x,y)}\right>\left<\psi_{(x,y)}\right|+\sum_{x,y}t\left|\psi_{(x,y)}\right>\left<\psi_{(x+1,y)}\right|\notag\\
&+&\sum_{x,y}t\left|\psi_{(x,y)}\right>\left<\psi_{(x,y+1)}\right| +H.c.
\label{H2c}
\end{eqnarray}
The bands of this model with periodical and open boundary condition are plotted in Fig. \ref{fig:4}b and c, with the number of sites in both $x$ and $y$ directions chosen to be 34, $t=-0.0051$
and $\lambda_x=\lambda_y=0.00741925$. For the value of $\lambda$ we use, $r_{(x,y)}$ varies from $0.081a$ to $0.108a$. By comparing those two figures, we can clearly observe the edge states appearing in the open boundary condition.  In Fig. \ref{fig:4}d we plot the wave function of the boundary state at $\phi_x=-2.443$, $\phi_y=0$ and $\omega_x=0.3427(2\pi c/a)$ using the tight binding basis at different sites $(x,y)$, which can be simply experimentally realized using the construction steps described in this Section.

%are band insulators with band inversion driven by spin-orbit coupling$^{5}$ and interaction$^{6}$.

\section{Methods}

\subsection{Topological Equivalence between 1D Models}
\begin{figure}[H]
\includegraphics[scale=.7]{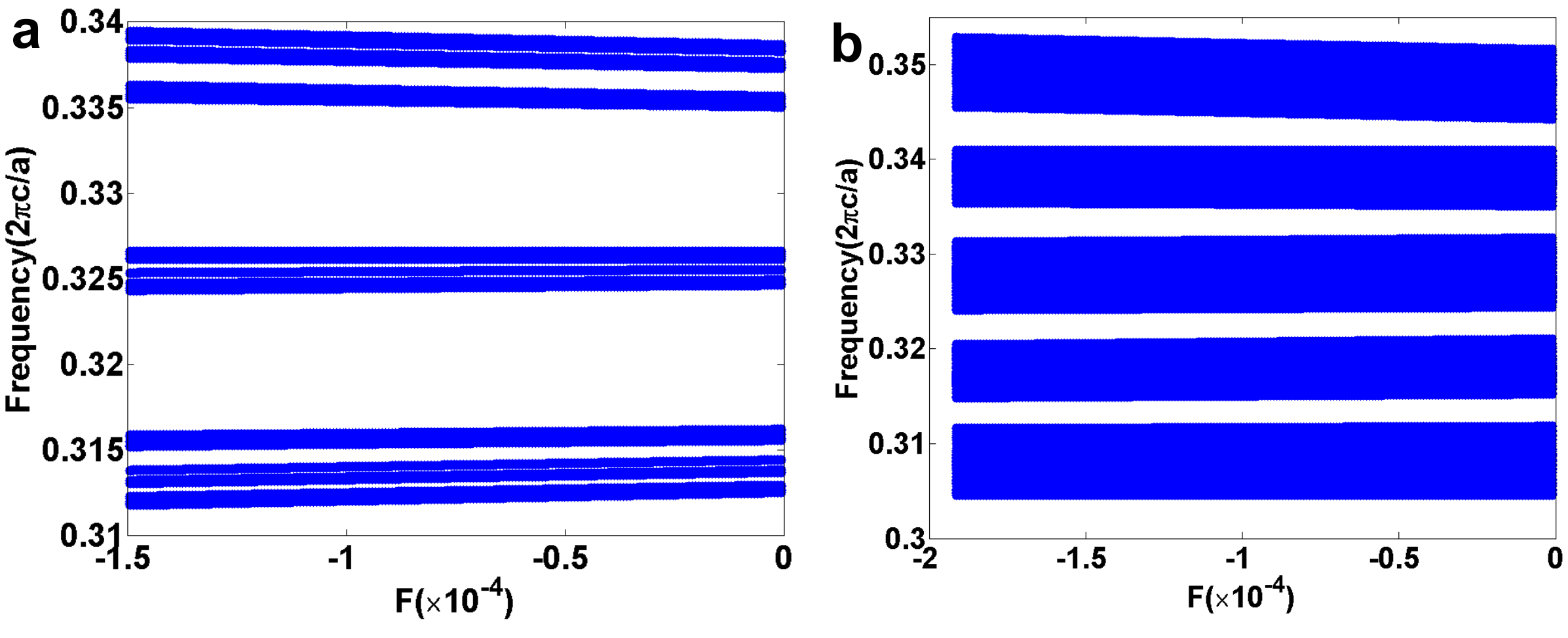}
\caption{(Color online) (a)The evolution of bulk bands of Eq .\ref{H1a} with first order approximation as a parameter of F defined in Eq. \ref{Va} and \ref{Wa}.(b) The evolution of bulk bands of Eq .\ref{H2a} with first order approximation as a parameter of F defined in Eq. \ref{V2a} and \ref{W2a}.   }
\label{fig:5}
\end{figure}

First we find that the hopping term in Eq. \ref{H1a} is a special case  of Eq. \ref{V} between sites $i$, $j$, so we just need to consider the generic case, defining $\Delta\bar{r}_{i,j}=\frac{r_i+r_j}{2}-r_0$, $\Delta r_{i,j}=\frac{r_i-r_j}{2}$, where $r_0$ is the average of the resonator radius $0.0965a$. Also we use first order approximation $r_x-r_0=c(\omega_x-\omega_{c0})=c\lambda_x\cos(2\pi bx+\phi_x)$, whereas $c$ is a constant
\begin{eqnarray}
V_{i,j}&=&V(r_i,r_j)\notag\\
&=&V(\Delta\bar{r}_{i,j}, \Delta r_{i,j})\notag\\
&=&V(0,0)+\frac{\partial V}{\partial \Delta\bar{r}_{i,j}}\Delta\bar{r}_{i,j}+\frac{\partial^2 V}{\partial \Delta r_{i,j}^2}(\Delta r_{i,j})^2\notag\\
&=&V(0,0)+\frac{\partial V}{\partial \Delta\bar{r}_{i,j}}\Delta\bar{r}_{i,j}+\frac{\partial^2 V}{\partial \Delta r_{i,j}^2}(\Delta r_{i,j})^2\notag\\
&\approx&V(0,0)+\frac{\partial V}{\partial \Delta\bar{r}_{i,j}}\frac{c\lambda_x}{2}(\cos(2\pi bi+\phi_x)+\cos(2\pi bj+\phi_x))+\frac{\partial^2 V}{\partial \Delta r_{i,j}^2}(\Delta r_{i,j})^2\notag\\
&\approx&t+\frac{\partial V}{\partial \Delta\bar{r}_{i,j}}\frac{c\lambda_x}{2}(\cos(2\pi bi+\phi_x)+\cos(2\pi bj+\phi_x))\notag\\
&=&t+F_1(\cos(2\pi bi+\phi_x)+\cos(2\pi bj+\phi_x))
\label{Va}
\end{eqnarray}
The first order term of $\Delta r_{i,j}$ vanishes because according to Eq. \ref{V}, $V_{i,j}$ is invariant under the exchange of $i,j$. In the second last step we ignored the second order term. Similarly
\begin{eqnarray}
\omega_x&=&\omega(r_x,\Delta r_{x,x-1},\Delta r_{x,x+1})\notag\\
&=&\omega(r_x,0,0)+\frac{\partial \omega}{\partial \Delta r_{x,x-1}}\Delta r_{x,x-1}+\frac{\partial \omega}{\partial \Delta r_{x,x+1}}\Delta r_{x,x+1}\notag\\
&=&\omega_x+c\frac{\partial \omega}{\partial \Delta r_{x,x-1}}\Delta \omega_{x,x-1}+c\frac{\partial \omega}{\partial \Delta r_{x,x+1}}\Delta \omega_{x,x+1}\notag\\
&=&\omega_x+F_2(\cos(2\pi bx+\phi_x)-\cos(2\pi b(x-1)+\phi_{x}))\notag\\
&+&F_2(\cos(2\pi bx+\phi_x)-\cos(2\pi b(x+1)+\phi_{x}))
\label{Wa}
\end{eqnarray}
According to numerical simulation, $F_1$ and $F_2$ are negative and positive and bounded by $F_m=0.0001492625\ll\left|t\right|,\left|\lambda\right|$ in Eq. \ref{H1c}. Thus Eq. \ref{H1a} can be very well approximated by Eq. \ref{H1c}. We can also see their topological equivalence by plotting the energy states at $\phi_x=0$ as a function of $F=F_1=-F_2=-\alpha F_m$, where $\alpha\in[0,1]$ as a control parameter (Fig. \ref{fig:5}a). Because the bulk band gaps did not close as $F$ evolves from $0$ to $-F_m$, the two models are topologically equivalent. 
\subsection{Topological Equivalence between 2D Models}
Same as the 1D case, the hopping term in Eq. \ref{H2a} is a special case  of Eq. \ref{V} between sites $i$, $j$, so we just need to consider the generic case, defining $\Delta\bar{r}_{i,j}=\frac{r_i+r_j}{2}-r_0$, $\Delta r_{i,j}=\frac{r_i-r_j}{2}$, where $r_0$ is the average of the resonator radius $0.095a$. Also we use first order approximation $r_{x,y}-r_0=c(\omega_{(x,y)}-\omega_{c0})=c(\lambda_x\cos(2\pi bx+\phi_x)+\lambda_y\cos(2\pi by+\phi_y))$, where c is a constant
\begin{eqnarray}
V_{i,j}&=&V(r_i,r_j)\notag\\
&=&V(\Delta\bar{r}_{i,j}, \Delta r_{i,j})\notag\\
&=&V(0,0)+\frac{\partial V}{\partial \Delta\bar{r}_{i,j}}\Delta\bar{r}_{i,j}+\frac{\partial^2 V}{\partial \Delta r_{i,j}^2}(\Delta r_{i,j})^2\notag\\
&=&V(0,0)+\frac{\partial V}{\partial \Delta\bar{r}_{i,j}}\Delta\bar{r}_{i,j}+\frac{\partial^2 V}{\partial \Delta r_{i,j}^2}(\Delta r_{i,j})^2\notag\\
&\approx&V(0,0)+\frac{\partial V}{\partial \Delta\bar{r}_{i,j}}\frac{c}{2}(\omega_i+\omega_j-2\omega_{c0})+\frac{\partial^2 V}{\partial \Delta r_{i,j}^2}(\Delta r_{i,j})^2\notag\\
&\approx&t+\frac{\partial V}{\partial \Delta\bar{r}_{i,j}}\frac{c}{2}(\omega_i+\omega_j-2\omega_{c0})
\label{V2a}
\end{eqnarray}
Similarly
\begin{eqnarray}
\omega_{(x,y)}&=&\omega(r_{(x,y)},\Delta r_{(x,y),(x-1,y)},\Delta r_{(x,y),(x+1,y)},\Delta r_{(x,y),(x,y-1)},\Delta r_{(x,y),(x,y+1)})\notag\\
&=&\omega(r_{(x,y)},0,0,0,0)+\frac{\partial \omega}{\partial \Delta r_{(x,y),(x-1,y)}}\Delta r_{(x,y),(x-1,y)}+\frac{\partial \omega}{\partial \Delta r_{(x,y),(x+1,y)}}\Delta r_{(x,y),(x+1,y)}\notag\\
&+&\frac{\partial \omega}{\partial \Delta r_{(x,y),(x,y-1)}}\Delta r_{(x,y),(x,y-1)}+\frac{\partial \omega}{\partial \Delta r_{(x,y),(x,y+1)}}\Delta r_{(x,y),(x,y+1)}\notag\\
&=&\omega_{(x,y)}+c\frac{\partial \omega}{\partial \Delta r_{(x,y),(x-1,y)}}\Delta \omega_{(x,y),(x-1,y)}+c\frac{\partial \omega}{\partial \Delta r_{(x,y),(x+1,y)}}\Delta \omega_{(x,y),(x+1,y)}\notag\\
&+&c\frac{\partial \omega}{\partial \Delta r_{(x,y),(x,y-1)}}\Delta \omega_{(x,y),(x,y-1)}+c\frac{\partial \omega}{\partial \Delta r_{(x,y),(x,y+1)}}\Delta \omega_{(x,y),(x,y+1)}
\label{W2a}
\end{eqnarray}
With $\lambda_x=\lambda_y$, we can define $F_1$ and $F_2$ the same as the 1D case. According to numerical simulation, $F_1$ and $F_2$ are negative and positive and bounded by $F_m=0.00019138\ll\left|t\right|,\left|\lambda\right|$ in Eq. \ref{H2c}. Thus Eq. \ref{H2a} can be very well approximated by Eq. \ref{H2c}. We can also see their topological equivalence by plotting the energy states at $\phi_x=\phi_y=0$ as a function of $F=F_1=-F_2=-\alpha F_m$, where $\alpha\in[0,1]$ as a control parameter (Fig. \ref{fig:5}b). Because the bulk band gaps did not close as $F$ evolves from $0$ to $-F_m$, the two models are topologically equivalent. 
\subsection*{References and Notes}
\begin{itemize}
\item[1.]
von Klitzing, Klaus The quantized Hall effect. {\it Rev. Mod. Phys.} {\bf 58}, 519-531 (1986).

\item[2.]
Qi, X. L. \& Zhang, S. C. The quantum spin Hall effect and
topological insulators. {\it Phys. Today} {\bf 63}, 33 (2010).
\item[3.]
Qi, X. L. \& Zhang, S. C. Topological insulators and
superconductors. {\it Rev. Mod. Phys.} {\bf 83}, 1057 (2011).
\item[4.]
Hasan, M. Z. \& Kane, C. L. Colloquium: Topological insulators. {\it
Rev. Mod. Phys.} {\bf 82}, 3045 (2010).

\item[5.]
Bernevig, B. A., Hughes, T. L. \& Zhang, S. C. Quantum Spin Hall
Effect and Topological Phase Transition in HgTe Quantum Wells. {\it
Science} {\bf 314}, 1757 (2006).
\item[6.]
Zhang, H.\&  Liu, C.-X. \& Qi, X.-L. \&  Dai, X. \&  Fang, Z. \&
Zhang, S.-C. Topological insulators in {Bi$_2$Se$_3$},
{Bi$_2$Te$_3$} and {Sb$_2$Te$_3$} with a single Dirac cone on the
surface. {\it Nat. Phys.} {\bf 5}, 438 (2009).
\item[7.]
Zhang, X. \& Zhang, H. J. \& Wang, J. \& Felser, C. \& Zhang, S.-C.
Actinide Topological Insulator Materials with Strong Interaction.
{\it Science} {\bf 335}, 1464 (2012).
\item[8.]
Chang, C. Z. {\it et al.}
Experimental Observation of the Quantum Anomalous Hall Effect in a Magnetic Topological Insulator.
{\it Science} {\bf 340}, 167 (2013).

\item[9.]
Yu, R. \& Zhang, W. \&  Zhang, H. J. \&  Zhang, S. C. \&  Dai, X. \&  Fang, Z.
Quantized Anomalous Hall Effect in Magnetic Topological Insulators.
{\it Science} {\bf 329}, 61 (2010).

\item[10.]
Zhang, X. \&  Wang, J. \& Zhang, S.-C. Topological insulators for
high-performance terahertz to infrared applications. {\it Phys. Rev.
B} {\bf 82}, 245107 (2010).

\item[11.]
Haldane, F. D. M. \& Raghu, S., Possible Realization of Directional Optical Waveguides in Photonic Crystals with Broken Time-Reversal Symmetry. {\it Phys. Rev. Lett.} {\bf 100}, 013904 (2008).

\item[12.]
Wang, Z. \& Chong, Y. D. \& Joannopoulos, J. D. \& Solja\ifmmode \check{c}\else \v{c}\fi{}i\ifmmode \acute{c}\else \'{c}\fi{}, M. Reflection-Free One-Way Edge Modes in a Gyromagnetic Photonic Crystal. {\it Phys. Rev. Lett.} {\bf 100}, 013905 (2008).

\item[13.]
Zhang, S.-C. \& Hu, J. P.
A Four-Dimensional Generalization of the Quantum Hall Effect.
{\it Science} {\bf 294}, 823-828 (2001).

\item[14.]
Kraus, Y. E. \& Ringel, Z. \&  Zilberberg, O.
Four-Dimensional Quantum Hall Effect in a Two-Dimensional Quasicrystal.
{\it Phys. Rev. Lett.} {\bf 111}, 226401 (2013).
\item[15.]
Kraus, Y. E. \& Lahini, Y. \& Ringel, Z. \& Verbin, M. \& Zilberberg, O.
Topological States and Adiabatic Pumping in Quasicrystals.
{\it Phys. Rev. Lett.} {\bf 109}, 106402 (2012).
\item[16.]
Verbin, M. \& Zilberberg, O. \& Kraus, Y. E. \& Lahini, Y. \& Silberberg, Y.
Observation of Topological Phase Transitions in Photonic Quasicrystals.
{\it Phys. Rev. Lett.} {\bf 110}, 076403 (2013).
\item[17.]
Kraus, Y. E. \& Zilberberg, O.
Topological Equivalence between the Fibonacci Quasicrystal and the Harper Model.
{\it Phys. Rev. Lett.} {\bf 109}, 116404 (2012).
\item[18.]
Villeneuve, P. R. \& Fan, S. H. \&  Joannopoulos, J. D.
Microcavities in photonic crystals: Mode symmetry, tunability, and coupling efficiency.
{\it Phys. Rev. B} {\bf 54}, 7837 (1996).
\item[19.]
Wang, Z. \& Fan, S. 
Magneto-optical defects in two-dimensional
photonic crystals.
{\it Appl. Phys. B} {\bf 81}, 369-375 (2005).
\item[20.]
Chien, F. S.-S  \&  Tu, J. B. \&  Hsieh, W.-F. \&  Cheng, S.-C.
Tight-binding theory for coupled photonic crystal waveguides.
{\it Phys. Rev. B} {\bf 75}, 125113 (2007).

\item[21.]
Mortensen N. A. \& Xiao S. \& Pedersen J. Liquid-infiltrated photonic crystals: enhanced light-matter interactions for lab-on-a-chip applications. {\it Microfluidics and Nanofluidics} {\bf 4}, 117-127 (2008).

\end{itemize}
\end{document}